\documentclass{article}

\usepackage{amssymb}
\usepackage{graphicx}
\begin{document}

\begin{center}

{\Large\bf FALSIFYING TREE LEVEL STRING\\[5PT]
MOTIVATED
BOUNCING COSMOLOGIES\\[5PT]}
\medskip

C.P. Constantinidis\footnote{e-mail: clisthen@cce.ufes.br},
J. C. Fabris\footnote{e-mail: fabris@cce.ufes.br}, R. G. Furtado\footnote{e-mail:
furtado@cce.ufes.br} \medskip

Departamento de F\'{\i}sica, Universidade Federal do Esp\'{\i}rito Santo,
29060-900, Vit\'oria, Esp\'{\i}rito Santo, Brazil
\medskip

N. Pinto-Neto\footnote{e-mail: nelsonpn@cbpf.br} and D. Gonzalez\footnote{e-mail:
diego@cbpf.br}
\medskip

ICRA - Centro Brasileiro de Pesquisas F\'{\i}sicas, CBPF. Rua Xavier Sigaud, 150,
Urca, CEP22290-180, Rio de Janeiro, Brasil

\end{center}

\begin{abstract}

The string effective action at tree level contains, in its bosonic
sector, the Einstein-Hilbert term, the dilaton, and the axion, besides
scalar and gauge fields coming from the Ramond-Ramond sector. The
reduction to four dimensions brings to scene moduli fields. We
generalize this effective action by introducing two arbitrary
parameters, $\omega$ and $m$, connected with the dilaton and axion
couplings. In this way, more general frameworks can be analyzed.
Regular solutions with a bounce can be obtained for a range of (negative)
values of the parameter $\omega$ which, however, exclude the pure
string configuration ($\omega = - 1$). We study the evolution of
scalar perturbations in such cosmological scenarios.
The predicted primordial power
spectrum decreases with the wavenumber with spectral index
$n_s=-2$, in contradiction with the
results of the $WMAP$. Hence, all such effective string motivated cosmological
bouncing models seem to be ruled out, at least at the tree level approximation.

\end{abstract}

Pacs numbers: 04.20.Cv., 04.20.Me,98.80.Cq

\section{Introduction}

The standard cosmological model, based on the general relativity
theory, has been successfully tested until the nucleosynthesis
era, around $t \sim 1\,s$, that is, at $T \sim
1\,MeV$\cite{kolb,bellido}. At times before this era, it is only
possible to speculate on how the universe has behaved. There are
some claims that the spectrum of the anisotropy of the cosmic
microwave background radiation (CMBR) has allowed to test the
inflationary scenario, opening a window to energy levels of about
$10^{15}\,GeV$\cite{riotto,dominik,kinney}. However, this last
statement remains controversial\cite{lue}: there is not, at this
moment, a unique, complete, inflationary model, free of problems
like transplanckian frequencies and fine tuning of fundamental
parameters. In this sense the inflationary model remains a
theoretical proposal asking for a full consistent formulation.
Moreover, the fitting of the CMBR spectrum is made
through a quite large number of parameters (up to ten), and more
results, like the identification of the gravitational waves
contribution to the CMBR anisotropy through the full detection of
the polarization parameters of the CMB photons are necessary in
order to surmount the degeneracy in the parameter space. Finally,
neither the standard cosmological model nor the inflationary
scenario address the question of the initial singularity, which is
a major problem in the primordial cosmology. Hence, the very early
universe remains an open area of research.
\par
String theory\cite{polchinski} is, at the moment, the most
important candidate to a fundamental theory of nature unifying all
interactions including gravity. At very low energy levels, string
theory reduces to general relativity in its gravity sector. On the other hand,
at higher energies, it predicts a deviation from the general
relativity framework: the dilaton field couples non-minimally to
the Einstein-Hilbert term; the Neveu-Schwarz sector exhibits also
an axion field, which in four dimensions is equivalent to a scalar
field which couples with the dilaton; in the Ramond-Ramond sector,
scalar fields and gauge fields, minimally coupled to gravity and
to the dilaton field, are present. Moreover, string theory is
formulated in ten dimensions, and the reduction to four dimensions
leads to the appearance of moduli fields associated with the
dynamics of the extra dimensions. String theory itself may be
incorporated in a more fundamental structure in $11$ dimensions
($M$ theory), $12$ dimensions ($F$ theory) and so on. All these
complex structures lead, in principle, to a strong deviation from
the standard cosmological scenario at very high energies.
\par
The question we address here is how such a rich
structure affects the evolution of the universe in its primordial
phase. This problem has already been studied in many aspects in
the literature\cite{lidsey}. The pre-big bang
scenario\cite{gasperini}, the ekpyrotic model\cite{ekp},
and the string gas cosmology \cite{brand1}
are some of them. However, all these
attempts are plagued with some important difficulties: to construct a
completely regular
cosmological scenario, without any kind of singularity, which consequences
can be successfully tested against observation.
\par
General scalar-tensor systems which reduce to the particular
string effective action for some values of the free parameters,
have been considered in the literature
\cite{picco,patrick,kirill}. In reference \cite{picco}, a $D$
dimensional lagrangian has been analyzed, with two free parameters
$\omega$ and $m$, which are connected with the coupling of the
dilaton and the axion fields, respectively. The original
configuration has been compactified considering flat, static extra
dimensions. For a specific range of values of $\omega$, scenarios
with no curvature singularity have been obtained. However, the dilaton
field is null initially, pointing out a
singularity in the string expansion parameter, and thus rendering the
effective model inadequate at that moment. In reference
\cite{patrick} fields from the Ramond-Ramond sector have been
considered. This allows to avoid the singularity in the string
expansion parameter, but only in the case of large negative values of $\omega$,
that is, $\omega < - 3/2$. This implies the presence of negative
energies when the lagrangian is re-expressed in the Einstein
frame. The $D$ dimensional structure has been studied, in the
vacuum case, in reference \cite{kirill} and regular solutions have
been obtained, again for large negative values $\omega$.
\par
In the present work, we perform a two-fold analysis. First, we
explore another possibility with respect to some previous works
(mainly references \cite{picco,patrick}): we take into account the
dynamics of the extra dimensions, and a radiative fluid whose
presence is suggested by a maxwellian field in the Ramond-Ramond
sector. Again, completely regular models can be obtained, this time for
a larger range of values of the parameter $\omega$, mainly with
respect to the results of reference \cite{patrick}: this parameter
must still be negative, but in four dimensions it can be greater
than $- 3/2$. The pure string case, which is characterized by
$\omega = - 1$, still leads to scenarios which are not completely
regular; in fact this particular value can imply regularity only
when the number of extra dimensions goes to infinity. On the other hand,
even if $\omega < - 1$, there are possible connections
between $M$ and $F$ theories. We describe the features of
these models, which display a bounce in the scale factor, while
the dilaton and the moduli fields remain regular. We then
test the new scenario obtained here against observations, as well as
those obtained in
\cite{picco,patrick}, by computing the
spectrum of scalar perturbations. The general result is that these
models predict a decreasing power spectrum,
which disagrees
strongly with observations\cite{teg}. As a general feature, all
the scenarios lead to a spectral index for scalar perturbations
that is around $-2$. This means that it is not possible to
construct a realistic regular cosmological scenario based on the
string (and more general frameworks) effective action at tree
level unless, perhaps, some non trivial compactification scheme is
introduced, or more complex configurations are considered like,
for example, the (condensate) fermionic terms.
\par
In the next section we describe the construction of the effective action.
In section $3$, the
background solutions are obtained. In section $4$ a perturbative analysis is
carried out
for the scalar modes and the power spectrum is computed. In section $5$ we
present our
conclusions.

\section{The effective action in four dimensions}

Our starting point is the string effective action in $D$
dimensions at tree level coupled to a radiative field:
\begin{equation}
\label{l1} L = \sqrt{- \tilde g}e^{-\sigma}\biggr\{\tilde R -
\tilde\omega\sigma_{;A}\sigma^{;A} -
\frac{1}{12}H_{ABC}H^{ABC}\biggl\} - \sqrt{- \tilde
g}\frac{\xi_{;A}\xi^{;A}}{2} + L_{matter} \quad ,
\end{equation}
where $A,B,C =1,...,D$, $\sigma$ is the dilatonic field, $H_{ABC}$
is the axionic field, $\xi$ is a scalar field coming from the
Ramond-Ramond sector and $L_{matter}$ represents the matter
lagrangian, in this case a radiative fluid. The origin of
the radiative fluid may be traced back to electromagnetic terms
present in the Ramond-Ramond sector. In the pure string case,
$\tilde\omega = - 1$, but other values of $\tilde\omega$ are
allowed if the effective action (\ref{l1}) is obtained from a
fundamental theory in higher dimensions, like M-theory or
F-theory. In these cases, the value of $\tilde\omega$ can be
different from $- 1$; in particular it can be largely negative
\cite{patrick}.
\par
The $D$-dimensional metric
describes now a manifold decomposition $M^{4}\times V^{D-4}$,
where $M^4$ is a four-dimensional space-time and $V^{D-4}$ is an
internal space which is supposed to be flat. The line element
takes the form
\begin{equation}
ds^2 = g_{\mu\nu}dx^\mu dx^\nu - e^{2\beta}dy^idy_i \quad ,
\end{equation}
where $\mu, \nu = 1,...4,$ and $i = 1,...,D-4$, with $\beta$
depending only on $x^{\mu}$. In this case, the
Ricci scalar in $D$ dimensions is
\begin{equation}
\tilde R = R - 2d\Box\beta - d(d + 1)\beta_{;\rho}\beta^{;\rho} \quad ,
\end{equation}
with $d = D - 4$, and $R$ being the scalar curvature in 4-dimensions.
The reduced lagrangian in four dimension reads,
\begin{eqnarray}
\label{l2} L &=& \sqrt{- g}e^{d\beta - \sigma}\biggr\{R -
2d\Box\beta - d(d + 1)\beta_{;\rho}\beta^{\rho} -
\tilde\omega\sigma_{;\rho}\sigma^{;\rho} -
\frac{1}{12}H_{\mu\nu\lambda}H^{\mu\nu\lambda}\biggl\}\nonumber\\
&-& \sqrt{- g}e^{d\beta}\frac{\xi_{;\rho}\xi^{;\rho}}{2} +
L_{matter} \quad ,
\end{eqnarray}
where we have supposed that the axion field $H_{ABC}$ and $\sigma$
also depend only on $x^{\mu}$. Defining $\phi = e^{d\beta -
\sigma}$, and performing integration by parts in the second term of
(\ref{l2}) the lagrangian reduces to
\begin{eqnarray}
\label{l3} L &=& \sqrt{- g}\phi\biggr\{R - \biggr[d(d + 1) +
\tilde\omega d^2\biggl]\beta_{;\rho}\beta^{\rho} -
\tilde\omega\frac{\phi_{;\rho}\phi^{;\rho}}{\phi^2} \nonumber \\
&+& 2d(1 + \tilde\omega)\beta_{;\rho}\frac{\phi^{;\rho}}{\phi} -
\phi^{-2}
\Psi_{;\rho}\Psi^{;\rho}\biggl\} \nonumber\\
&-& \sqrt{- g}e^{d\beta}\frac{\xi_{;\rho}\xi^{;\rho}}{2} +
L_{matter} \quad .
\end{eqnarray}
The axion field has been written as $H^{\mu\nu\nu} =
\phi^{-1}\epsilon^{\mu\nu\lambda\gamma}\Psi_{;\gamma}$. The final
form of the lagrangian is
\begin{eqnarray}
\label{l4} L &=& \sqrt{- g}\phi\biggr\{R +
\gamma_{;\rho}\gamma^{;\rho}
-\omega\frac{\phi_{;\rho}\phi^{;\rho}}{\phi^2} -
\phi^{m-1}\Psi_{;\rho}\Psi^{;\rho}\biggl\}\nonumber\\
&-& \sqrt{-
g}\phi^{-\frac{ds}{r}}e^{\frac{d}{r}\gamma}\frac{\xi_{;\rho}\xi^{;\rho}}{2}+
L_{matter} \quad ,
\end{eqnarray}
where we have defined $\gamma = r\beta + s\ln\phi$ with $r^2 = -
[d(d + 1) + \tilde\omega d^2]$ and $s = d(1 + \tilde\omega)/r$.
The parameter $\omega$ is given by
\begin{equation}
\omega = - \frac{(d - 1)\tilde\omega + d}{d(1 + \tilde\omega) + 1}
\quad .
\end{equation}
In this effective lagrangian, we have introduced an arbitrary
constant $m$ in the coupling between the dilaton and axion field.
In the pure string case, $m = - 1$ and $\omega = -1$. However, in what
follows, we will keep $m$ and $\omega$ arbitrary in order to have
contact with general multidimensional and supergravity theories,
as well as some extensions of the string theory, like $M$ or $F$
theories. Notice that $\tilde\omega = - 1$ implies $\omega = - 1$.
Hence, the traditional string configuration is a kind of fixed
point. The form of the lagrangian (\ref{l4}) is valid under the
condition $\tilde\omega < - (d + 1)/d$ ($r$ is real). This excludes the strict
string case, but it includes $\tilde\omega > - 3/2$. The consequences
of supposing $\tilde\omega > - (d + 1)/d$ (which amounts to change
the sign of the term in $\gamma$) will be discussed later.
\par
From (\ref{l4}), the field equations read:
\begin{eqnarray}
\label{fe1}
 R_{\mu\nu} - \frac{1}{2}g_{\mu\nu}R &=&
\frac{8\pi}{\phi}T_{\mu\nu} +
\frac{\omega}{\phi^2}\biggr(\phi_{;\mu}\phi_{;\nu} -
\frac{1}{2}g_{\mu\nu}\phi_{;\rho}\phi^{;\rho}\biggl) \nonumber\\
&+& \frac{1}{\phi}\biggr(\phi_{;\mu;\nu} -
g_{\mu\nu}\Box\phi\biggl) - \biggr(\gamma_{;\mu}\gamma_{;\nu} -
\frac{1}{2}g_{\mu\nu}\gamma_{;\rho}\gamma^{;\rho}\biggl)
\nonumber\\&+&
\phi^{m - 1}\biggr(\Psi_{;\mu}\Psi_{;\nu} -
\frac{1}{2}g_{\mu\nu}\Psi_{;\rho}\Psi^{;\rho}\biggl) \nonumber\\
&+& \frac{\phi^{-\frac{ds}{r} -
1}e^{\frac{d}{r}\gamma}}{2}\biggr(\xi_{;\mu}\xi_{;\nu} -
\frac{1}{2}g_{\mu\nu}\xi_{;\rho}\xi^{;\rho}\biggl) \quad ; \\
\label{fe2} \Box\phi + \frac{1 - m}{3 +
2\omega}\phi^m\Psi_{;\rho}\Psi^{;\rho} &+& \frac{1}{3 +
2\omega}\biggr(\frac{ds}{r} +
1\biggl)\phi^{-\frac{ds}{r}}e^{\frac{d}{r}\gamma}
\frac{\xi_{;\rho}\xi^{;\rho}}{2} \nonumber\\
&=& \frac{8\pi}{3 + 2\omega}T \quad ; \\
\label{fe3}
\Box\Psi + m\frac{\phi^{;\rho}}{\phi}\Psi_{;\rho} &=& 0  \quad ;\\
\label{fe4} \Box\gamma + \frac{\phi^{;\rho}}{\phi}\gamma_{;\rho} +
\frac{1}{4}\frac{d}{r}\phi^{- \frac{ds}{r} - 1}
e^{\frac{d}{r}\gamma}\xi_{;\rho}\xi^{;\rho} &=& 0 \quad ;\\
\label{fe6} \Box\xi -
\frac{ds}{r}\frac{\phi_{;\rho}\xi^{;\rho}}{\phi} +
\frac{d}{r}\gamma_{;\rho}\xi^{;\rho} &=& 0 \quad ; \\
\label{fe5} {T^{\mu\nu}}_{;\mu} &=& 0 \quad .
\end{eqnarray}
The energy-momentum tensor has the perfect fluid form $T^{\mu\nu} = (\rho + p)u^\mu u^\nu -
pg^{\mu\nu}$.

\section{Regular cosmological solutions}

Let us consider a homogeneous and isotropic space-time described
by the
Friedmann-Robertson-Walker metric
\begin{equation}
ds^2 = dt^2 - a^2(t)\biggr\{\frac{dr^2}{1 - kr^2} + r^2(d\theta^2 +
\sin^2\theta d\phi^2)\biggl\} .
\end{equation}
As usual, $a(t)$ is the scale factor, and $k = -1, 0, 1$
represents the normalized curvature of the spatial section.
Inserting the metric in the field equations, we
obtain the following equations of motion:
\begin{eqnarray}
\label{em1}
3\biggr(\frac{\dot a}{a}\biggl)^2 + 3\frac{k}{a^2} &=& \frac{8\pi}{\phi}\rho +
\frac{\omega}{2}\biggr(\frac{\dot\phi}{\phi}\biggl)^2 - 3\frac{\dot a}{a}\frac{\dot\phi}{\phi}
- \frac{\dot\gamma^2}{2} + \frac{\phi^{m - 1}}{2}\dot\Psi^2 \nonumber\\
&+& \frac{\phi^{-\frac{ds}{r} - 1}e^{\frac{d}{r}\gamma}}{4}\dot\xi^2 \quad ,\\
\label{em2} \ddot\phi + 3\frac{\dot a}{a}\dot\phi + \frac{1 - m}{3
+ 2\omega}\phi^m\dot\Psi^2 &+& \frac{1}{3 +
2\omega}\biggr(\frac{ds}{r} +
1\biggl)\phi^{-\frac{ds}{r}}e^{\frac{d}{r}\gamma}
\frac{\dot\xi^2}{2} \nonumber\\&=& \frac{8\pi}{3 + 2\omega}(\rho - 3p) \quad ,\\
\label{em3}
\ddot\Psi + 3\frac{\dot a}{a}\dot\Psi + m\frac{\dot\phi}{\phi}\dot\Psi &=& 0 \quad ,\\
\label{em4} \ddot\gamma + 3\frac{\dot a}{a}\dot\gamma +
\frac{\dot\phi}{\phi}\dot\gamma + \frac{1}{4}\frac{d}{r}\phi^{-
\frac{ds}{r} - 1}e^{\frac{d}{r}\gamma}\dot\xi^2 &=& 0
\quad ,\\
\label{em5} \ddot\xi + 3\frac{\dot a}{a}\dot\xi -
\frac{ds}{r}\frac{\dot\phi}{\phi}\dot\xi +
\frac{d}{r}\dot\gamma\dot\xi &=& 0 ,\\
 \label{em6}
\dot\rho + 3\frac{\dot a}{a}(\rho + p) &=& 0 \quad .
\end{eqnarray}
It will be supposed that the ordinary fluid obeys a barotoropic equation of state,
$p = \alpha\rho$, with $-1 \leq \alpha \leq 1$.
\par
Equations (\ref{em3},\ref{em5},\ref{em6}) admit the first
integrals
\begin{equation}
\dot\Psi = A\frac{\phi^{-m}}{a^3} \quad ,  \quad \dot\xi =
\xi_0\frac{\phi^{\frac{ds}{r}}e^{- \frac{d}{r}\gamma}}{a^3} \quad
, \quad \rho = \rho_0a^{-3(1 + \alpha)} \quad ,
\end{equation}
where $A$, $\xi_0$ and $\rho_0$ are integration constants. In the
absence of the RR field $\xi$, equation (\ref{em4}) admits also a
first integral given by
\begin{equation}
\dot\gamma = \frac{B}{a^3\phi} \quad ,
\end{equation}
where $B$ is a constant

In the absence of the moduli field, the equation for
the RR field is obtained from Eq.~(\ref{l3}) by making $\beta=$const.
and absorving $\exp (d\beta)$ in the definition of $\xi$,
yielding $\Box\xi=0$ with solution $\dot\xi=\xi_0/a^3$.
\par
It seems that there exists no general closed solution when all
fields are taken into account. However, particular cases can be
explicitly solved. The solutions without the moduli fields were
found in reference \cite{patrick}; the main scenarios for this
case will be reviewed below. The equations, in the presence of moduli
fields, but in the absence of the Ramond-Ramond (RR) term, will be
solved for two main cases: in the absence of the axion field, and
in the presence of the axion field. Only the flat case will be
treated. The final conclusions may also be applied to the non flat
cases.

\subsection{Solutions in absence of the moduli fields}

When the internal scale factor is static, regular solutions, with
no divergence in the dilaton field and the space-time curvature, are
obtained both in the absence of the RR scalar field and in the
presence of the RR scalar field. For both cases, the coupling
constant $\omega$ must be smaller than $- 3/2$ in order to obtain
a completely regular scenario. This means that in the Einstein
frame, negative energies are present. The flat solutions of
interest were obtained in reference \cite{patrick}. They are the
following:
\begin{itemize}
\item Pure anomolous axionic case ($\omega < - 3/2, m = - 1$):
\begin{eqnarray}
\label{aa1}
a(\tau) &=& \frac{a_0}{\sqrt{\sinh\alpha\tau}}\biggr[1 - \sin
f(\tau)\biggl]^{-1/2} \quad , \\
\phi(\tau) &=& \phi_0\sinh(\alpha\tau) \quad , \\
\Psi' &=& A\phi \quad , \\
f(\tau) &=&
\ln\biggr[\biggr|\tanh\biggr(\frac{\alpha\tau}{2}\biggl)\biggl|^p\biggl]
\quad ,
\end{eqnarray}
where
\begin{eqnarray}
\label{aa2}
\alpha = \sqrt{\frac{- 2A^2}{3 + 2\omega}}
\quad , \quad p = \pm \sqrt{-\biggr(1 + \frac{2}{3}\omega\biggl)} \quad ,
\nonumber\\ a_0^2 = \frac{A^2\phi_0}{3M}\quad ,
\quad M=\frac{8\pi\rho_0 G}{3c^2}
\end{eqnarray}
where $A$ is a constant. The time parameter $\tau$ is related to
the cosmic time $t$ by the relation $dt = a^3d\tau$
and a prime means derivative with respect to $\tau$.
\item Anomolous RR case ($\omega < - 3/2, m = - 1$):
\begin{eqnarray}
a(\tau) &=& \frac{a_0}{\sqrt{\sinh(\alpha\tau) - s_0}}\biggr[1 -
\sin f(\tau)\biggl]^{-1/2} \quad , \\
\phi &=& \phi_0[\sinh(\alpha\tau) - s_0]\quad , \\
\Psi '&=& A\phi \quad , \\
\xi &=& \chi_0 = {\rm const.} \quad , \\
f(\tau) &=&
p\ln\biggr|\frac{2}{s_0}\biggr[\frac{s_0\tanh(\alpha\tau/2) + 1 -
\sqrt{1 + s_0^2}}{1 + \sqrt{1 + s_0^2} +
s_0\tanh(\alpha\tau/2)}\biggl]\biggl| \quad ,
\end{eqnarray}
\end{itemize}
where now
\begin{equation}
a_0^2 = \frac{A^2\phi_0}{3M}(1+s_0^2)\quad {\rm and}
\quad s_0=\frac{\xi_0^2}{2A^2\phi_0} > 1.
\end{equation}

In the two cases there are bounces in the scale factor.
One can be obtained by choosing $\tau_i$ and $\tau_f$
such that $f(\tau_i)=-7\pi/2$ and $f(\tau_f)=-3\pi/2$.
It represents an asymptotically radiation dominated universe contracting
from infinity which bounces at $\tau_B$ such that
$f(\tau_B)=-5\pi/2$, expanding aterwards to another asymptotically
radiation dominated phase. The dilaton is always finite,
going from $\phi_i/\phi_0=10^{-1}$ to $\phi_f/\phi_0=10^{2}$,
implying a reduction of the effective gravitational constant.
In this sense, such solutions represent a completely regular
scenario (see Ref.\cite{patrick} for details).
However, they are plagued by the presence of negative energies in
the Einstein frame, which may lead to instabilities at the quantum
level, even if the stability at classical level is
assured\footnote{Since our physical frame is the Jordan's frame,
the question of quantum stability is much more delicate.}.
Solutions where this problem is ameliorated occur if the
moduli fields is taken into account, as it will be done in the next
subsection.

\subsection{Solutions with moduli and without the axion and RR fields}

Let us set the RR and the axion fields as constants, that is, $A =
0$ and $\xi_0 = 0$. In this case, the field equations can be
easily solved if the new time coordinate $\tau$ is employed as
before. With this time re-parametrization, equation (\ref{em2})
admits a simple solution:
\begin{equation}
\phi = \phi_0\tau \quad , \quad \phi_0 = \mbox{constant} \quad .
\end{equation}
To solve the einsteinian equation (\ref{em1}), it is better to
reexpress it in the Einstein frame. This is achieved by writing
$a = \phi^{-1/2}b$. Hence, equation (\ref{em1}) reduces to
\begin{equation}
\biggr(\frac{b'}{b}\biggl)^2 = \frac{\tilde{M}b^2 - E}{\tau^2} \quad ,
\quad \tilde{M} = \frac{8\pi G\rho_0}{3c^2\phi_0^2} \quad , \quad E =
\frac{B^2}{6\phi^2_0} - \frac{3 + 2\omega}{12} \quad .
\end{equation}
The prime means derivative with respect to the new time
coordinate $\tau$. When $E > 0$, the solution for the original
scale factor $a(\tau)$ is
\begin{equation}
\label{sf1} a(\tau) =
\frac{1}{\sqrt{\tau}}\frac{a_0}{\cos[\sqrt{E}\ln\tau]} \quad ,
\quad a_0 = \mbox{constant} \quad .
\end{equation}
\par
The solution described by (\ref{sf1}) represents a non-singular
universe, which exhibits a bounce. At same time the "effective
gravitational coupling" represented by the inverse of the field
$\phi$ is also regular, evolving from a finite value to another
finite value. The scale factor of the internal dimension is given
by the expression,
\begin{equation}
e^\beta = e^\frac{\gamma}{r}\phi^\frac{-s}{r} \quad .
\end{equation}
With the choice of the minus sign in the definition of $r$, the
solutions found before imply a decreasing internal scale factor,
with finite values at the beginning and at the end, that is, the
internal scale factor stabilizes. On the other hand, the string
expansion parameter is given by $g_s = e^{\tilde\sigma} =
e^{d\beta}/\phi$, and it is also a decreasing function, remaining
always finite. In this sense, the effective action at tree level
remains always meaningful. Due to the absence of singular
behaviour in all three relevant quantities, the solutions found
above represent a regular cosmological model. The behaviour of
these functions, together with the effective gravitational
coupling, is displayed in figure $1$.
\begin{figure}
\begin{center}
\includegraphics[width=7cm]{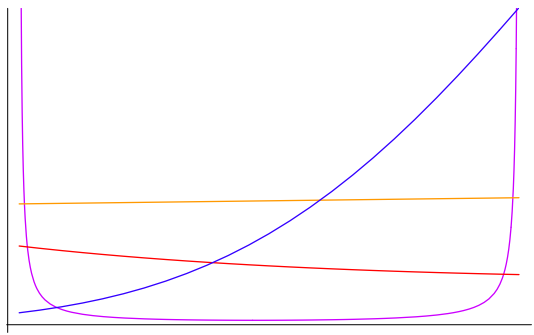}
\end{center}
\caption{Behaviour of the scale factor (violet), internal scale
factor (red), effective gravitational coupling (blue) and the
string expansion parameter (yellow).}
\end{figure}
\par
The case $\omega > - (d + 1)/d$, implying $E < 0$, which includes
the pure string configuration, has been treated in reference
\cite{shinji}, for an arbitrary dimension $D$. It is equivalent to
changing the sign of the kinetic term for the field $\gamma$ in
(\ref{l3}). The solution exhibits singularities, especially in the
dilaton field: such singularity implies a breakdown of the string
expansion, and the lagrangian (\ref{l3}) is not valid anymore.

\subsection{Solutions with the moduli and axion fields, without the RR field}

Let us now consider, besides the moduli fields, the axion field,
that is $A \neq 0$, still in the absence of the RR field, that is,
$\xi_0 = 0$. Equation (\ref{em2}) admits a simple solution if we
choose a new time coordinate such that
\begin{equation}
dt = \tilde{E}\phi^\frac{1 + m}{2}a^3d\vartheta \quad , \quad \tilde{E} =
\frac{2}{1 - m}\sqrt{\frac{3 + 2\omega}{2A^2}} \quad ,
\end{equation}
that is,
\begin{equation}
\phi = \phi_0\sin^\frac{2}{1 - m}\vartheta \quad , \quad \phi_0 =
\biggr\{\frac{(3 + 2\omega)C}{2A^2}\biggl\}^{\frac{1}{1 - m}}
\quad,
\end{equation}
$C$ being an integration constant.
In terms of this new time parameter, and after the redefinition
$a = \phi^{-1/2}b$, equation
(\ref{em1}) takes the form, for the flat case,
\begin{equation}
\bigg(\frac{b'}{b}\biggl)^2 = \frac{3 + 2\omega}{3(1 -
m)^2}\biggr\{\frac{1}{\bar C}\biggr[{M}b^2 - \frac{B^2}{6}\biggl]+
1\biggl\}\frac{1}{\sin^2\vartheta} \quad ,
\end{equation}
where a prime means now derivative with respect to $\vartheta$, and
\begin{equation}
\bar C =\frac{3 + 2\omega}{12}C \quad \mbox{and} \quad {M} =
\frac{8\pi G\rho_0}{3c^2} \quad,
\end{equation}
leading to the relation
\begin{equation}
\label{ie} \int\frac{db}{b\sqrt{\tilde{M}b^2 - N^2}} = \pm \frac{1}{1 -
m}\sqrt{\frac{3 + 2\omega}{3\bar C}} \ln\tan\frac{\vartheta}{2}
\quad ,
\end{equation}
where $N^2 = B^2/6 - \bar C$. Notice that this excludes the case
$\bar C > B^2/6$ since it would imply a non-regular solution, at
least in the dilatonic field, as the final results indicate.
\par
The result for the scale factor is:
\begin{equation}
a(\vartheta) = a_0\sin^{-1/(1 -
m)}\vartheta\frac{1}{\cos[pf(\vartheta)]} \quad;
\end{equation}
where
\begin{equation}
\quad f(\vartheta) = \ln\tan\frac{\vartheta}{2}\ \quad , \quad p =
\pm \frac{2}{1 - m}\frac{N}{\sqrt{C}} \quad ,
\end{equation}
and
\begin{equation}
a_0 = \biggl(\frac{6\bar C}{A^2}\biggr)^{1/[2(m-1)]}\frac{N}{\sqrt{\tilde{M}}} .
\end{equation}
Moreover,
\begin{eqnarray}
\Psi' &=& \Psi_0\sin\xi \quad , \\
\gamma' &=& \frac{\gamma_0}{\sin\vartheta} \quad ,
\end{eqnarray}
where
\begin{equation}
\Psi_0 = \frac{(3+2w)\sqrt{C}}{A(1 - m)},
\end{equation}
\begin{equation}
\gamma_0 = \frac{2B}{(1 - m)\sqrt{C}} .
\end{equation}
\par
The solutions found above describe a completely regular cosmological
scenario. The scale factor behaves asymptotically as $a \propto
|t|^\frac{1}{2}$ for $t \rightarrow \pm \infty$. The internal
scale factor and the dilatonic field (consequently its inverse,
the string perturbative parameter) $e^{-\sigma}$ are given by
\begin{eqnarray}
e^\beta =
\biggr[\tan\frac{\vartheta}{2}\biggl]^\frac{\gamma_0}{r}\phi(\vartheta)^\frac{-s}{r}
\quad , \\
e^{-\sigma} =
\biggr[\tan\frac{\vartheta}{2}\biggl]^\frac{-d\gamma_0}{r}\phi(\vartheta)^\frac{r
+ ds}{r}  \quad .
\end{eqnarray}
The solution is completely regular in the sense that there is no
curvature singularity. Also, the effective gravitational
coupling $\phi$, the internal scale factor $e^\beta$ and the
dilatonic field $e^{-\sigma}$ behave regularly. Moreover, the
internal dimensions become small and stabilize for some choices of
the integration parameters. The behaviour of these functions are
similar to those displayed in figure $1$.
\par
In order for having a bounce in the flat case in the Einstein frame,
which is the case of the solutions presented above, the null
energy condition $\rho + p
> 0$ must be violated near the bounce \cite{patrick27}, being satisfied
only far from the bounce. The fields which violate
the null energy condition can be easily seen from the Lagrangians
in the Einstein frame:
\begin{equation}
\label{e14a} L = \sqrt{- g}\biggr\{R +
\gamma_{;\rho}\gamma^{;\rho} -(\omega +
3/2)\chi_{;\rho}\chi^{;\rho} - \exp
(-2\chi)\Psi_{;\rho}\Psi^{;\rho}\biggl\} + L_{matter}  ,
\end{equation}
when the moduli field is present, and
\begin{eqnarray}
\label{e15a} L &=& \sqrt{- g}\biggr\{R + -(\omega +
3/2)\chi_{;\rho}\chi^{;\rho} - \exp
(-2\chi)\Psi_{;\rho}\Psi^{;\rho} - \exp
(-\chi)\xi_{;\rho}\xi^{;\rho}\biggl\}\nonumber\\
&+& L_{matter}  .
\end{eqnarray}
In the first case, the moduli field is the one which violates
the null energy condition, while in the second case, if
$\omega < -3/2$, is the dilaton which violate this condition:
both have the ``wrong'' sign in their kinetic terms.

\section{Spectrum of scalar perturbations}

Before writing the perturbed equations, let us redefine the scalar fields as follows:
\begin{eqnarray}
\chi &=& \ln\phi \quad , \\
\Psi &=& \zeta \quad . \\
\end{eqnarray}
All the fields are made a-dimensional. Hence, the field $\phi$ is
redefined as \cite{weinberg}
\begin{equation}
\frac{3 + 2\omega}{4 + 2\omega}G\phi \quad \rightarrow\quad \phi
\quad .
\end{equation}
Moreover, the quantities with dimension of time or space are made
dimensionless by using the Planck's time ($t_{Pl}$) or length
($L_{Pl}$). The scale factor has no dimension. In what follows,
the perturbative analysis will be performed in the Einstein frame,
for technical simplicity. The final spectrum is computed in the
Jordan frame.
\par
The metric, including the background and the perturbed functions,
has the form
\begin{equation}
ds^2 = b^2(\eta)[(1 + 2\Phi)d\eta^2 - (1 - 2\Psi)\gamma_{ij}dx^idx^j]
\quad ,
\end{equation}
where $\Phi = \Phi(\eta,\vec x)$ and $\Psi = \Psi(\eta,\vec x)$
are the metric fluctuations in the longitudinal gauge, $\eta$ is
the conformal time, $dt = bd\eta$, and $b$ is the scale factor in
the Einstein frame. We follow here the gauge invariant formalism
\cite{brand}. We also define ${\cal{H}} = b_{\eta}/b$ (from now on, subscripts
$\eta$ indicate derivatives with respect to the conformal time $\eta$).
It is convenient to write separately the final perturbed equations
for the case where the RR field is present and the moduli fields
are absent and for the case where the RR field is absent and the
moduli fields are present.
\begin{itemize}
\item Perturbed equations in the absence of the RR field:
\begin{eqnarray}
\label{pe1}& & \Phi_{\eta\eta} + 4{\cal{H}}\Phi_{\eta} - \biggr\{4k +
\frac{\nabla^2}{3}\biggl\}\Phi =
\frac{\bar\omega}{3}\chi_{\eta}\delta\chi_{\eta}\nonumber\\
&+& \frac{m - 1}{6}\zeta_{\eta}^2e^{(m - 1)\chi}\delta\chi -
\frac{\gamma_{\eta}}{3}\delta\gamma_{\eta} + \frac{\zeta_{\eta}}{3}e^{(m -
1)\chi}\delta\zeta_{\eta}
\quad ,
\\
\label{pe2} & &\delta\chi_{\eta\eta} + 2{\cal{H}}\delta\chi_{\eta}
- \biggr\{\nabla^2 + \frac{(1 - m)^2}{2\bar\omega}e^{(m -
1)\chi}\zeta_{\eta}^2\biggl\}\delta\chi -
4\chi_{\eta}\Phi_{\eta}\nonumber\\
&+& \frac{(1 - m)}{\bar\omega}e^{(m - 1)\chi}
\zeta_{\eta}\delta\zeta_{\eta} = 0 \quad ,\\
\label{pe3} & &\delta\zeta_{\eta\eta} + [2{\cal{H}} + (m -
1)\chi_{\eta}]\delta\zeta_{\eta} - \nabla^2\delta\zeta -
4\Phi_{\eta}\zeta_{\eta}\nonumber\\
&+&
(m - 1)\zeta_{\eta}\delta\chi_{\eta} = 0 \quad ,\\
\label{pe4} & &\delta\gamma_{\eta\eta} + 2{\cal{H}}\delta\gamma_{\eta} -
\nabla^2\delta\gamma - 4\Phi_{\eta}\gamma_{\eta} = 0 \quad ,
\end{eqnarray}
where
\begin{equation}
\bar\omega = 3/2 + \omega \quad . \end{equation}
\item Perturbed
equations in the absence of the moduli fields ($m = - 1$):
\begin{eqnarray}
\label{pe1bis} \Phi_{\eta\eta} + 4{\cal{H}}\Phi_{\eta} - \biggr\{4k +
\frac{\nabla^2}{3}\biggl\}\Phi =
\frac{\bar\omega}{3}\chi_{\eta}\delta\chi_{\eta} & &\nonumber\\
+ \frac{e^{-2\chi}}{3}\zeta_{\eta}\delta\zeta_{\eta} -
\biggr[\frac{e^{-2\chi}}{3}\zeta_{\eta}^2 +
\frac{e^{-\chi}}{12}\xi_{\eta}^2\biggl]\delta\chi +
\frac{e^{-\chi}}{6}\xi_{\eta}\delta\xi_{\eta}& &
\quad ,\\
\label{pe2bis} \delta\chi_{\eta\eta} + 2{\cal{H}}\delta\chi_{\eta} -
\biggr\{\nabla^2 +
\frac{1}{4\bar\omega}e^{-\chi}\xi_{\eta}^2 +
\frac{2}{\bar\omega}e^{-2\chi}\zeta_{\eta}^2\biggl\}\delta\chi
& &\nonumber\\
- 4\chi_{\eta}\Phi_{\eta}+ \frac{2}{\bar\omega}e^{-2\chi}\zeta_{\eta}
\delta\zeta_{\eta} +
\frac{1}{2\bar\omega}e^{- \chi}\xi_{\eta}\delta\xi_{\eta}&=& 0 \quad ,\\
\\
\label{pe3bis} \delta\zeta_{\eta\eta} + 2({\cal{H}} - \chi_{\eta})
\delta\zeta_{\eta} -
\nabla^2\delta\zeta - 4\Phi_{\eta}\zeta_{\eta} - 2\zeta_{\eta}\delta\chi_{\eta}
&=& 0 \quad , \\
\nonumber\\
\label{pe4bis} \delta\xi_{\eta\eta} + (2{\cal{H}} - \chi_{\eta})\delta\xi_{\eta}-
\nabla^2\delta\xi - 4\Phi_{\eta}\xi_{\eta}- \xi_{\eta}\delta\chi_{\eta}&=& 0
\quad ,
\end{eqnarray}
where now
\begin{equation}
\bar\omega = 3/2 + \tilde\omega \quad .
\end{equation}
\end{itemize}
In equations (\ref{pe1}-\ref{pe4}) and (\ref{pe1bis}-\ref{pe4bis})
the perturbed variables are $\Phi$, $\delta\chi$, $\delta\zeta$,
$\delta\gamma$ and $\delta\xi$, and the background quantities are
${\cal{H}}$, $\chi$, $\gamma$, $\zeta$ and $\xi$.
\par
Equations (\ref{pe1}-\ref{pe4}) and (\ref{pe1bis}-\ref{pe4bis})
are written using the conformal time. The background solutions are
expressed in terms of the time parameter $\tau$ (for $m = - 1$) or
$\vartheta$ (for $m \neq - 1$). If we consider the perturbed
equations (\ref{pe1bis}-\ref{pe4bis}), for example, and changing
to the time parameter $\tau$, we obtain,
\begin{eqnarray}
\label{pe1bisbis} \Phi'' + (2h + \chi')\Phi' - \biggr\{4k
+ \frac{\nabla^2}{3}\biggl\}e^{-2\chi}b^4\Phi =
\frac{\bar\omega}{3}\chi'\delta\chi' & &\nonumber\\
+ \frac{e^{-2\chi}}{3}\zeta'\delta\zeta' -
\biggr[\frac{e^{-2\chi}}{3}\zeta'^2 +
\frac{e^{-\chi}}{12}\xi'^2\biggl]\delta\chi +
\frac{e^{-\chi}}{6}\xi'\delta\xi'& &
\quad , \\
\nonumber\\
\label{pe2bisbis} \delta\chi'' + \chi'\delta\chi' -
\biggr\{e^{-2\chi}b^4\nabla^2 +
\frac{1}{4\bar\omega}e^{-\chi}\xi'^2 +
\frac{2}{\bar\omega}e^{-2\chi}\zeta'^2\biggl\}\delta\chi
& &\nonumber\\
- 4\chi'\Phi'+
\frac{2}{\bar\omega}e^{-2\chi}\zeta'\delta\zeta' +
\frac{1}{2\bar\omega}e^{- \chi}\xi'\delta\xi'&=& 0 \quad ,\\
\nonumber\\
\label{pe3bisbis} \delta\zeta'' - \chi'\delta\zeta '-
e^{-2\chi}b^4\nabla^2\delta\zeta - 4\Phi'\zeta' - 2\zeta'\delta\chi' &=& 0
\quad , \\
\nonumber\\
\label{pe4bisbis} \delta\xi'' - e^{-2\chi}b^4\nabla^2\delta\xi
- 4\Phi'\xi'- \xi'\delta\chi'&=& 0 \quad ,
\end{eqnarray}
where $h = b'/b$. These are the equations which will be
integrated numerically. The other cases follow the same lines.
\par
The operator
$\nabla^2$ acts on the three-dimensional section at constant
curvature, and its eigenfunctions are plane waves ($k = 0$),
spherical harmonics ($k = 1$) and pseudo-spherical harmonics ($k =
- 1$). From now on, we will restrict ourselves to the flat case.
Hence, the perturbed quantities, represented generically by
$\delta_n$, can be decomposed into Fourier components:
\begin{equation}
\delta(\eta,\vec x) = \frac{1}{(2\pi)^{3/2}}\int
\delta_n(\eta,\vec n)e^{i\vec n.\vec x}d^3n \quad ,
\end{equation}
where the Fourier components $\delta_k$ obey the Helmholtz
equation
\begin{equation}
\nabla^2\delta_n = - n^2\delta_n \quad .
\end{equation}
We are adopting the convention that the coordinates have
dimensions of length while the metric is dimensionless.
Hence the wave number has dimensions of ${\rm cm}^{-1}$.
The range of interest for the
wavenumber $\vec n$ corresponds to scales where
the approximation of isotropy and homogeneity is valid. Today,
these scales are roughly between $100\,Mpc$ and $3.000\,Mpc$. In
order to fix the initial interval of $\vec n$ corresponding to
those scales today, we match asymptotically the solutions found
with the standard cosmological phase. This task becomes easier since
the asymptotic behaviour of the solutions presented here coincides
with the radiative phase of the standard cosmological model. If
$\tau_f$ ($\vartheta_f$) corresponds to the final asymptote of the
solutions, we fix this matching at $\tau_m$ ($\vartheta_m$) such
that
\begin{equation}
\frac{\tau_f - \tau_m}{\tau_m} = 10^{-5} \quad .
\end{equation}
The choice of the precision of the matching does not affect
essentially the final results unless it is too small, so that
numerical computation may become doubtful.
\par
From $\tau = \tau_m$ ($\vartheta = \vartheta_m$) on the Universe
follows the evolution dictated by the standard cosmological model.
We generally impose that this happens before the nucleosynthesis
period. Hence, imposing that the scale factor today equals unity
(such that today the physical scales correspond to the coordinate
scales), this normalization implies that the matching must occur
when $a < 10^{-12}$. Consider, for example, that the matching
is made exactly at $a = 10^{-12}$, and that at this moment the
gravitational coupling is constant, so that the formulation in the
Einstein frame becomes identical to that of the Jordan frame, $a=b$.

To be specific, let us take the pure anomalous axionic solution of
subsection (3.1) as an example (the other solutions follow the same lines),
with $\omega=-3$. We define the dimensioless parameter $\theta=\alpha\tau$
with range $2{\rm arctgh}\exp (-7\pi/2)<\theta<2{\rm arctgh}\exp (-3\pi/2)$.
In $\theta_m$ we have
\begin{eqnarray}
b(\theta_m) &=& b_0 \biggr[1 - \sin f(\theta)\biggl]^{-1/2} = 10^{-12}
\quad , \\
\phi(\theta_m) &=& \phi_0\sinh(\theta_m) = 1 \quad ,
\end{eqnarray}
remembering that, for this $\omega$,
\begin{equation}
f(\theta) =\ln\biggr[\biggr|\tanh\biggr(\frac{\theta}{2}\biggl)\biggl|\biggl].
\end{equation}
These relations lead to $b_0 = 10^{-18}$ and $\phi_0=\sinh(3\pi/2)$.
Using Eqs~(\ref{aa2}), we obtain
\begin{equation}
\alpha = A\sqrt{\frac{2}{3}} = 10^{-50}{\rm cm}^{-1},
\quad , \quad p = 1 \quad {\rm and} \quad M=10^{-60}{\rm cm}^{-2} .
\end{equation}
Equations (\ref{pe1bisbis}--\ref{pe4bisbis}) then reads
\begin{eqnarray}
\label{pe1bisbis2} & & \Phi_{\theta\theta} + \biggl\{\frac{\cos
f(\theta)}{[1-\sin f(\theta)]\sinh(\theta)} + {\rm
cotgh}(\theta)\biggr\} \Phi_{\theta}\nonumber\\
&+& \frac{{\tilde{n}}^2}{3\sinh^2(\theta)[1-\sin
f(\theta)]^2\sinh^2(3\pi/2)}\Phi \nonumber \\&=& -{\rm
cotgh}(\theta)\delta\chi_{\theta} +
\frac{\delta\zeta_{\theta}}{\sqrt{6}\sinh(3\pi/2)\sinh(\theta)} -
\frac{3}{2}\delta\chi \quad ,
\\
& &\delta\chi_{\theta\theta} + {\rm cotgh}(\theta)\delta\chi_{\theta} +
\biggr\{\frac{{\tilde{n}}^2}{\sinh^2(\theta)[1-\sin f(\theta)]^2\sinh^2(3\pi/2)}
-1\biggl\}\delta\chi \nonumber\\ &=&
4{\rm cotgh}(\theta)\Phi_{\theta} +
\frac{\sqrt{2}}{\sqrt{3}\sinh(3\pi/2)\sinh(\theta)}\delta\zeta_{\theta}
\quad ,
\\
& &\delta\zeta_{\theta\theta} - {\rm cotgh}(\theta)\delta\zeta _{\theta}+
\frac{{\tilde{n}}^2}{\sinh^2(\theta)[1-\sin f(\theta)]^2\sinh^2(3\pi/2)}
\delta\zeta \nonumber\\ &=& 2\sqrt{6}\sinh(3\pi/2)\sinh(\theta)\Phi_{\theta}
+\sqrt{6}\sinh(3\pi/2)\sinh(\theta) \delta\chi_{\theta}
\quad .
\end{eqnarray}
In these equations $\tilde n \equiv nb_0^2/\alpha$.
\par
The final step is to fix the initial conditions. The most
natural way is to impose that the initial spectrum is determined
by quantum fluctuations. The determination of this spectrum
follows the standard procedure\cite{brand,nelson}. First of all we
must remark that the perturbed functions decouple for $\eta
\rightarrow - \infty$. Hence, all perturbed quantities behave as
free fields. The metric fluctuation must be expressed in terms of
a new variable that is formally equivalent to a free scalar field.
At this moment, the fields are quantized, leading to normalized
modes. Choosing an initial vacuum state, the initial spectrum
expressed in the conformal time reads

\begin{eqnarray}
\Phi_n \propto 3\sqrt{\frac{3}{2}}\frac{\beta}{{\cal{H}}n^2}
\biggr(\frac{v_n}{z}\biggl)_{\eta}
\quad ,\\
v_n = \frac{1}{\sqrt{2c_sn}}e^{-ic_s n\eta} \quad ,
\end{eqnarray}
where
$\beta = {\cal{H}}^2 - {\cal {H}}'$, $z = \frac{b\beta^{1/2}}{c_sH}$
and $c_s = 1/\sqrt{3}$.
For the other fields, represented generically by $u$, one has
\begin{equation}
u_n \propto \frac{1}{\sqrt{2n}}e^{-i n\eta} \quad .
\end{equation}
In terms of the $\theta$ parameter one has:
\begin{eqnarray}
\label{psiini}
\Phi_n^{\rm ini}&\propto&\frac{3^{3/4}\sinh^3(7\pi/2)
(\theta-\theta_i)^2}{8b_0^3{\tilde{n}}^{3/2}}\biggl[\frac{\sqrt{3}
(\theta_i-\theta)\sinh^2(7\pi/2)}{4\tilde{n}b_0^2}
+i\biggr]\nonumber\\ &\times&\exp\biggl[\frac{i\tilde{n}4b_0^2}
{\sqrt{3}\sinh^2(7\pi/2)(\theta-\theta_i)}\biggr]
\end{eqnarray}
and
\begin{equation}
\label{fieldsini} \delta\chi_k^{\rm ini}\propto\frac{\sinh(7\pi/2)
(\theta-\theta_i)}{2b_0{\tilde{n}}^{1/2}}
\exp\biggl[\frac{i\tilde{n}4b_0^2}
{\sinh^2(7\pi/2)(\theta-\theta_i)}\biggr] .
\end{equation}
\par
With these initial conditions, the system is left to evolve. Due
to the complexity of the perturbed equations and of the background
solutions, the system of equations is solved numerically, with the
initial conditions fixed as described before, using the software
{\it Mathematica}. The processed final spectrum is given by
\begin{equation}
P_n = {\bar\Phi}_n^2\,n^3 \propto n^{n_s}\quad,
\end{equation}
where $n_s$ is the spectral index. The Bardeen potential
$\bar\Phi$ in the Jordan frame is
related to the potential in the Einstein frame by
\begin{equation}
\bar\Phi = \Phi + \frac{\delta\chi}{2} \quad .
\end{equation}
The spectrum is computed at the moment the universe matches the
standard model in the radiative era before the nucleosynthesis.
The general behaviour of the spectrum, for all cases, is given in
figure $1$ for a large range of values of $n$. In general, it
display an oscillating behaviour with decreasing amplitude.
However, the cosmological scales of interest, determined by the
matching conditions together with the normalization of the scale
factor today, imply $n << 1$. For this case, and for scales
today between $100\,Mpc$ and $3.000\,Mpc$ the spectrum has the
shape displayed in figure $2$, again for all the cases treated in
the present study. It is a decreasing spectrum, in strong
disagreement with the observation which indicates a nearly
Harrisson-Zeldovich spectrum \cite{bennett}. It is remarkable that
in all different cases the spectral index is essentially $n = -
2$. We have computed the spectrum at constant time and, in this
situation, a Harrison-Zeldovich spectrum implies that $P_n$ grows
with $n$.
\begin{figure}
\begin{center}
\includegraphics[width=7cm]{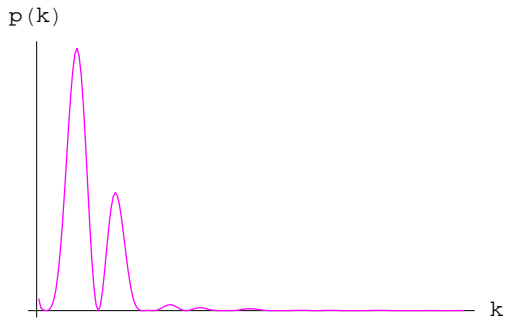}
\end{center}
\caption{General form of the spectrum.}
\end{figure}
\begin{figure}
\begin{center}
\includegraphics[width=7cm]{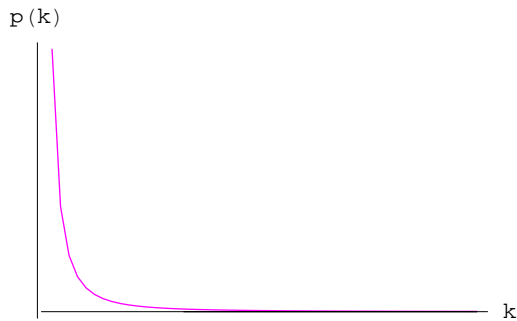}
\end{center}
\caption{The predicted spectrum for scales up to the Hubble
radius.}
\end{figure}

\section{Conclusions}

We have studied cosmological models based on the string
effective action at tree level. The dilaton, the axion, RR and the
moduli fields were taken into account, besides gauge fields coming
from the Ramond-Ramond sector. Two free parameters were introduced
in the effective lagrangian, $\omega$ and $m$, connected with
the coupling of the dilaton and axion fields. These parameters
were kept free in order to cover fundamental theories, like $M$
and $F$ theories, as well as $p$-branes configurations in the
superstring theory. Regular solutions were found, but only for
cases where $\omega < - 1$, which excludes the pure string case.
The new solutions found here, taking into account the moduli
fields, exhibit compactification of the extra dimensions. The
scale factor displays a bounce, exhibiting initially a contracting
phase before entering the expansion phase. Asymptotically, a
radiative phase is recovered allowing to match the solution with the standard
cosmological model. At the same time, the string expansion parameter
remains finite during the whole evolution of the universe. The effective
gravitational coupling decreases with time what, at least,
alleviate the hierarchial problem related to the characteristic
energy scale of gravity.
\par
The properties of the solutions found reveal that such "string"
cosmological models can be candidates for describing the
primordial phase of the universe. However, traces from this
primordial phase can be compared with observations through the
evaluation of the power spectrum of scalar fluctuations. This
primordial power spectrum is inferred from the spectrum of the
anisotropy of the CMBR. The observational results favor a flat
spectrum. Here, we have computed the spectrum at the beginning of
the radiative phase supposing that the initial fluctuations were
formed in the beginning of the contraction phase, before the
bounce, and that they were of quantum mechanical origin. The
quantum fluctuations are compatible with the asymptotical
behaviour in the beginning of the contracting phase. However, the
final spectrum is strongly decreasing, in contradiction with the
observational results, which favors at least a quasi-scale
invariant spectrum. Hence, the regular solutions found, in spite
of their nice features, are not candidate for a realistic
primordial cosmological model, The same features are found for the
regular "anomalous" solutions found in reference \cite{patrick}.

In fact, the result $n_s=-2$ for the spectral index is not
surprising. In reference \cite{nelson}, the same spectrum has been
obtained in the Einstein frame for a scalar field with negative
kinetic energy together with radiation. Here the situation is very
similar, see equations (\ref{e14a},\ref{e15a}), but with the
presence of other fields. These other fields do not alter the
bounce itself and, most important, the asymptotic behaviour. It
seems that the intermediate phase they can be important is not
relevant for the spectrum at large scales. Going to the Jordan's
frame does not modify the result because the spectrum of
$\delta\chi$ is negligible with respect to that of $\Phi$ for
small wave numbers.
\par
These results seem to exclude such string motivated models at tree level.
Nevertheless, it must be stressed that in the model developed here a quite
simple compactification mechanism was considered: the internal
space is flat, a $d$-dimensional torus. String theory admits many
other kinds of compactifications. In particular, the Calabi-Yau
manifolds are especially interesting since they can accommodate, in
quite natural way, the gauge groups of the standard model of
particle physics \cite{green}. In this case, the effective model
in four dimensions will be different to the one we have studied
here. However, the negative results reported here indicate
difficulties in constructing meaningful realistic cosmological models based
on string motivated effective actions at tree level.

\vspace{1.0cm}

{\bf Acknowledgments:} We thank CNPq (Brazil) and CAPES (Brazil)
for partial financial support. We thank also J\'er\^ome Martin and
Patrick Peter for their criticisms and suggestions.

\end{document}